\documentclass[12pt]{article}
\usepackage{graphicx}
\begin{document}

\bigskip
\centerline{\bf Optimization of hierarchical structures of information flow}

\bigskip

\noindent
D. Stauffer$^1$ and P.M.C. de Oliveira$^2$

\medskip
\noindent
Laboratoire PMMH, \'Ecole Sup\'erieure de Physique et de Chimie
Industrielles, 10 rue Vauquelin, F-75231 Paris, France

\bigskip
\noindent
$^1$ Visiting from:

\noindent
Inst. for Theoretical Physics, Cologne University, D-50923 K\"oln, Euroland

\medskip
\noindent
$^2$ Visiting from:

\noindent
Instituto de F\'{\i}sica, Universidade
Federal Fluminense; Av. Litor\^{a}nea s/n, Boa Viagem,
Niter\'{o}i 24210-340, RJ, Brazil


\bigskip
{\bf Abstract}

{\small The efficiency of a large hierarchical organisation is simulated 
on Barab\'asi-Albert networks, when each needed link leads to a loss
of information. The optimum is found at a finite network size, corresponding
to about five hierarchical layers, provided a cost for building the network
is included in our optimization.
} 

\section{Introduction}

In a hospital, the information on how to treat a patient has to travel from
the experts to the nurses, and errors may occur in this process. They can be 
avoided if only the leading expert deals with the patients, but then only
few patients can be treated. In the opposite extreme one builds in many layers
of hierarchy where the information has to be given from each layer to the 
larger layer below it until the nurses get the information. For a large number 
of layers, growing in size from top to bottom, one then treats badly a huge
number of patients. What is the optimum number of layers, or more generally,
the optimum organisational structure, for the hospital?

Analogous problems occur elsewhere in society \cite{braha}. Leaders 
(government, university president, company CEO) think that they know best, 
but we minor subjects not always fully appreciate their wisdom since something 
got lost in translation from one level to the lower level. 

We assume the loss of information from one layer to the next to be a fixed 
fraction $x$ either of the initial information (linear decay) or of the current 
information (exponential decay). If the top layer has index $L=0$, then at 
layer $L$ of the original information only the fraction exp($-xL$) or $1-xL$
arrives, for exponential or linear decay, respectively.

In a directed square lattice such information flow was already treated in an
old Citation Classic \cite{pmco}. In the next section, we deal with it on 
a Cayley tree analytically, while in section 3 we simulate it on directed 
Barab\'asi-Albert networks \cite{BA}.

\begin{figure}[hbt]
\begin{center}
\includegraphics[angle=-90,scale=0.5]{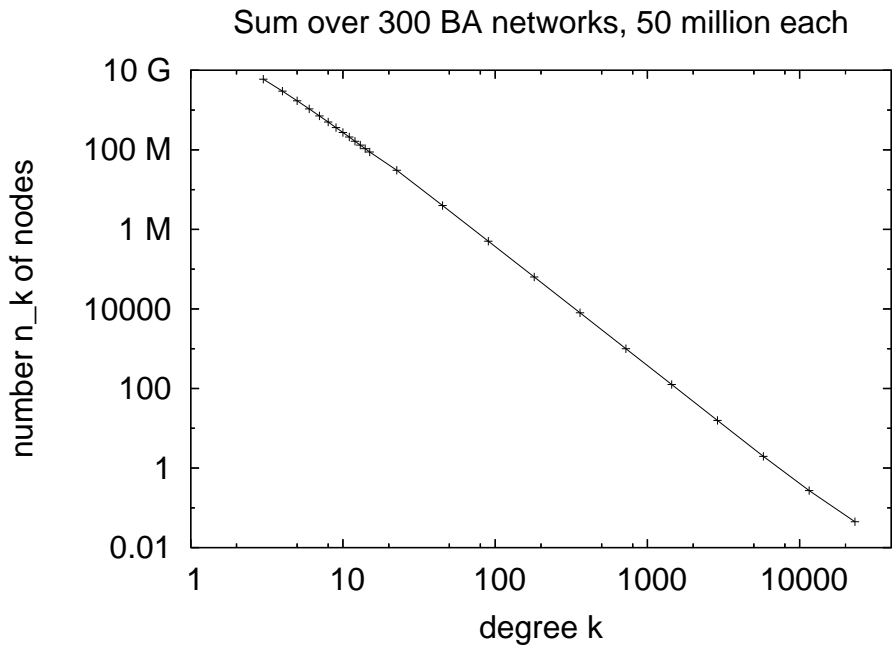}
\end{center}
\caption{Number $n_k$ of nodes having $k$ neighbours each, summed over
300 Barab\'asi-Albert networks with $m = 3$ and 50 million nodes each.
}
\end{figure}

\begin{figure}[hbt]
\begin{center}
\includegraphics[angle=-90,scale=0.31]{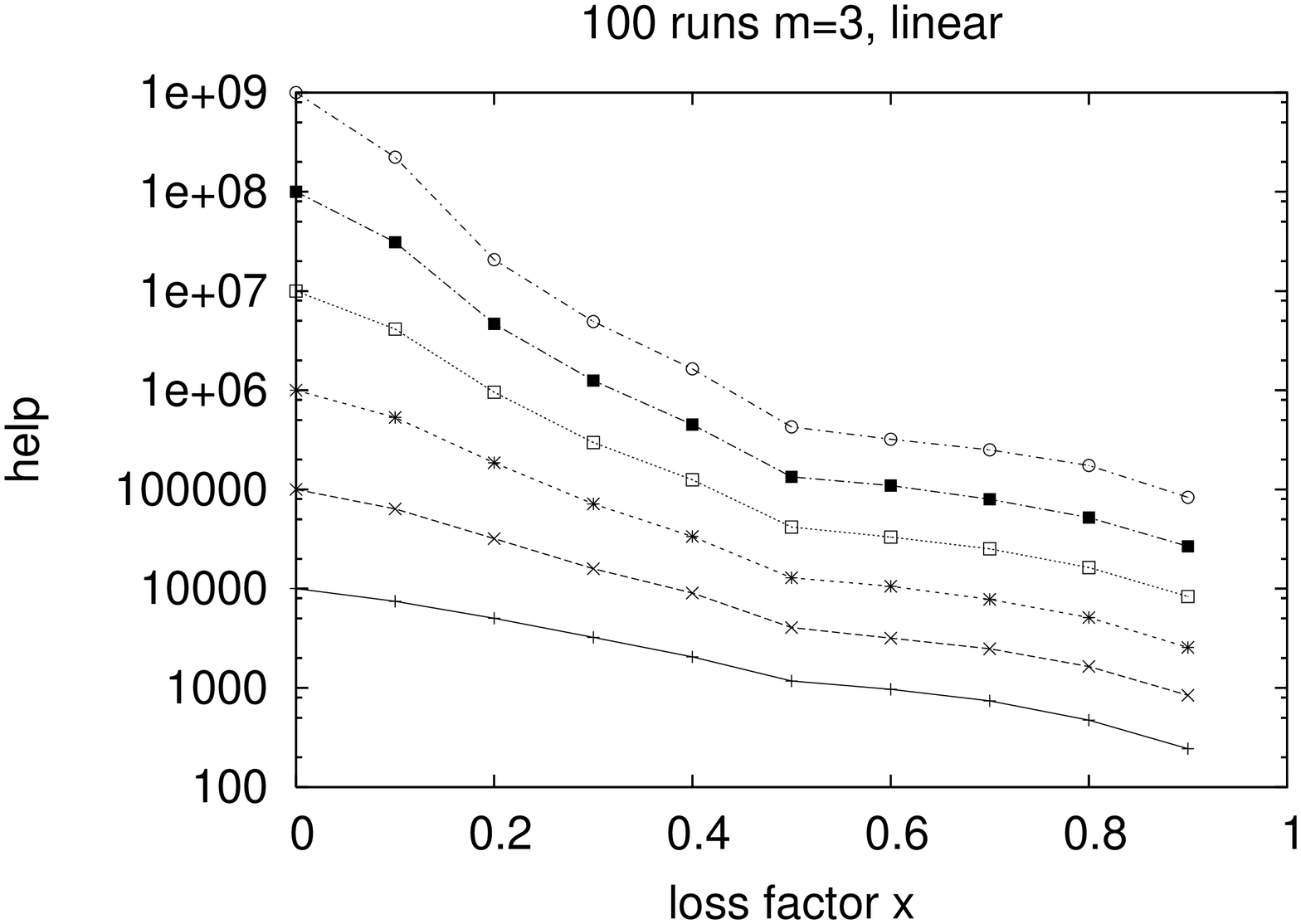}
\includegraphics[angle=-90,scale=0.31]{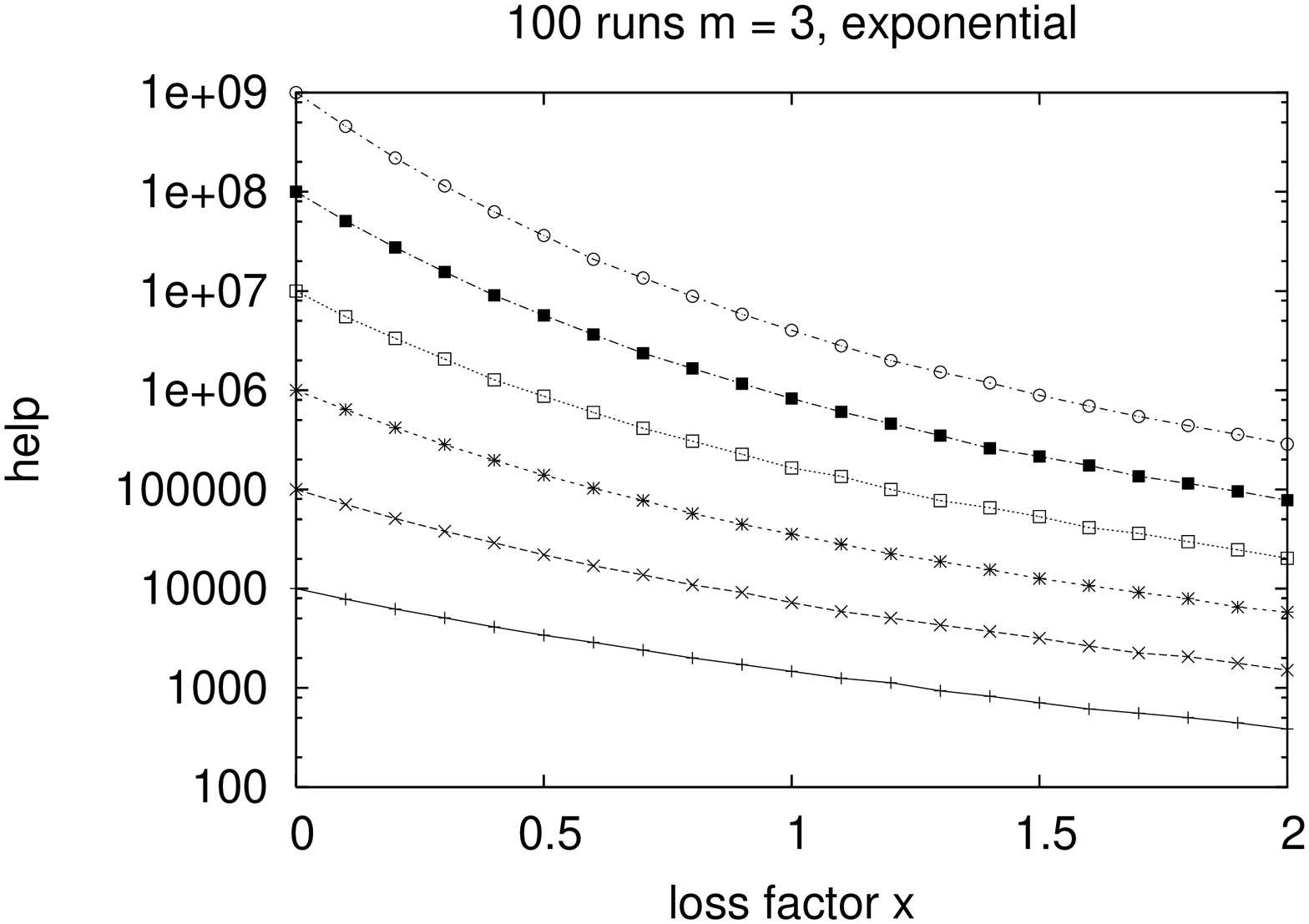}
\end{center}
\caption{Help summed over 100 scale-free networks of $N = 10^2, \, 10^3, \, 
10^4, \, 10^5, \, 10^6, \, 10^7$ nodes each, with $m=3$, versus loss
fraction $x$ and linear (top) or exponential (bottom) 
information decay.
}
\end{figure}

\begin{figure}[hbt]
\begin{center}
\includegraphics[angle=-90,scale=0.31]{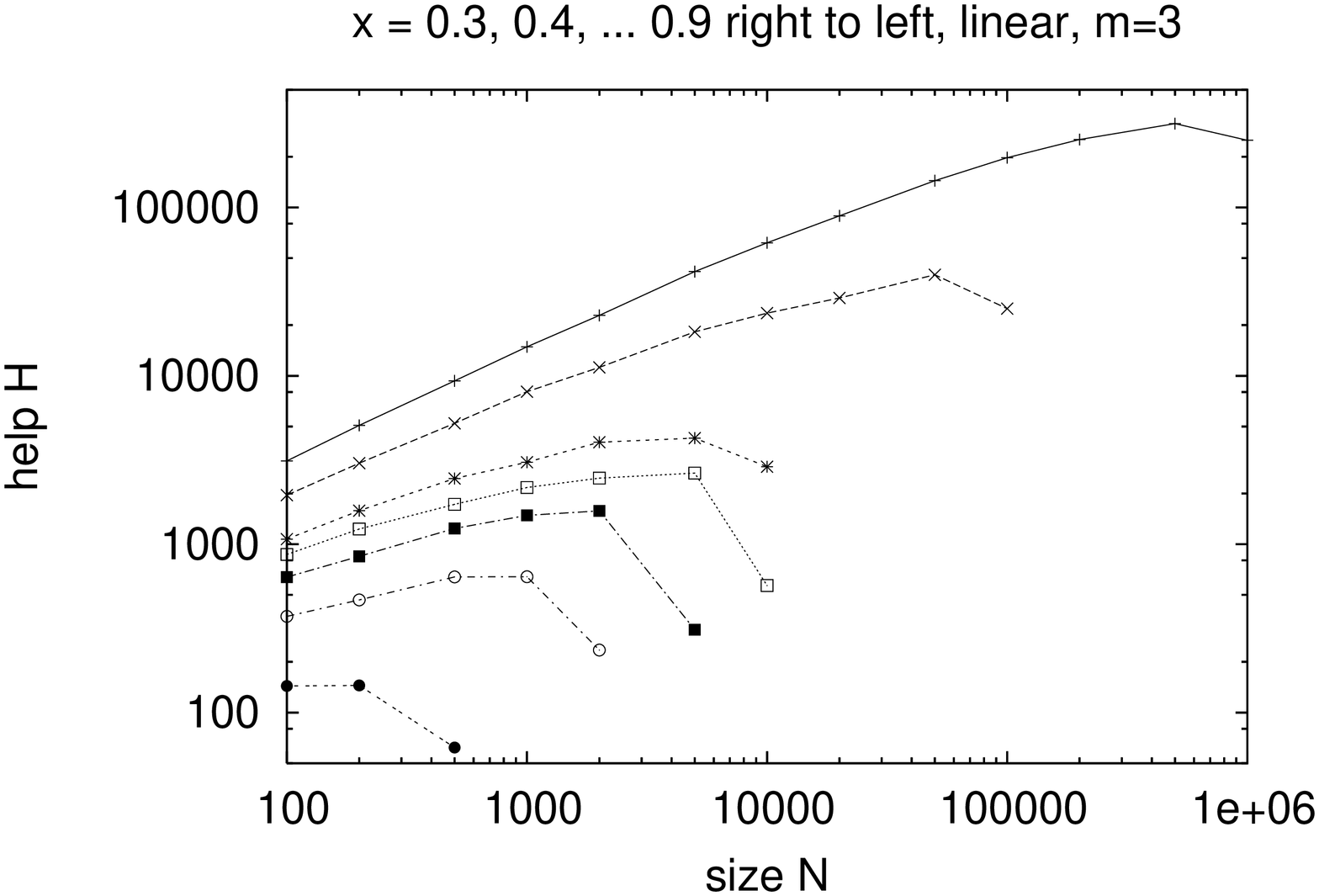}
\includegraphics[angle=-90,scale=0.31]{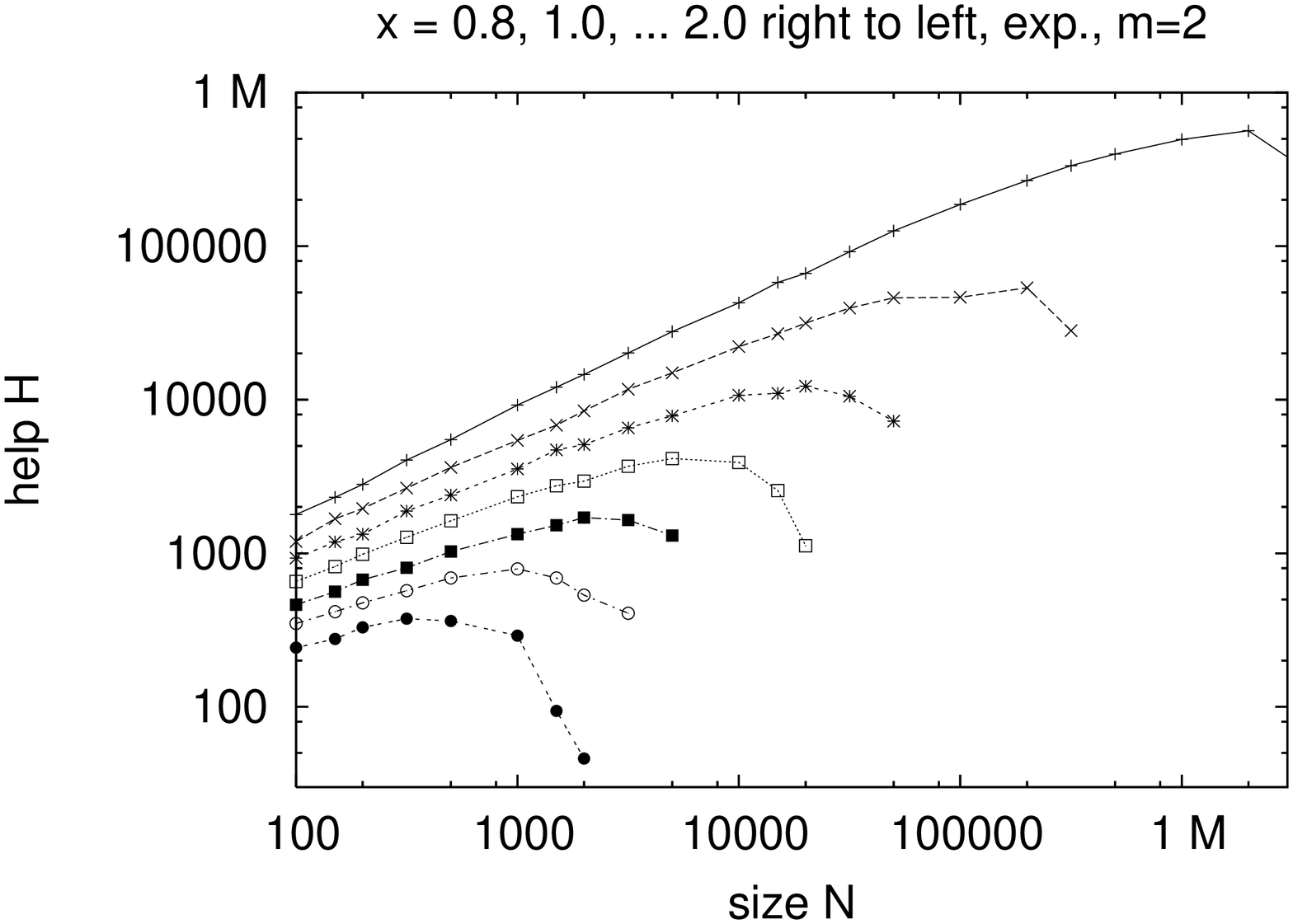}
\end{center}
\caption{Help minus size/100, summed over 100 samples, for linear and 
exponential decay.
}
\end{figure}

\begin{figure}[hbt]
\begin{center}
\includegraphics[angle=-90,scale=0.5]{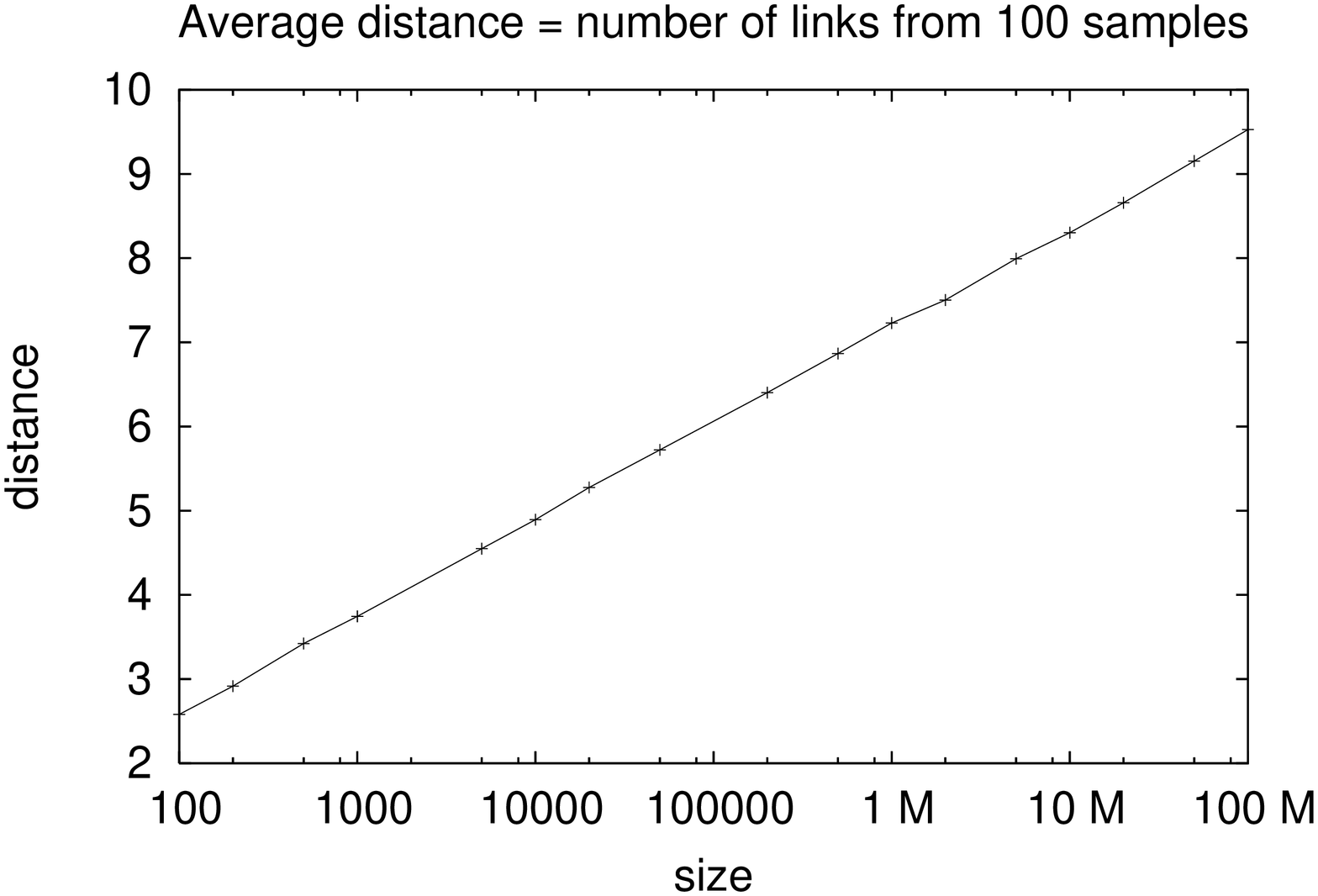}
\end{center}
\caption{Number of layers, defined as the average distance $\ell$ from the
core, versus number $N$ of nodes in 100 Barab\'asi-Albert networks with $m=2$.
}
\end{figure}

\section{Cayley tree}

Imagine there is one omniscient expert on top in layer 0, who talks to $b$ 
subordinates in layer 1, each of which again talks to $b$ different 
subordinates in layer 2, etc. With $L$ layers below the top we have in total
$$N = (b^{L+1}-1)/(b-1) \eqno (1)$$
people, and the bottom layer has a distance of $L$ from the top (as measured by 
the number of connecting links.) This Cayley tree or Bethe lattice is well
known to be analytically solvable in many applications. We measure the help
$H$ (or profit, or utility function) by the total amount of information 
arriving in the bottom layer. This help is
$$ H = (1-xL)b^L  \quad {\rm linear,\; for} \; L < 1/x  \eqno(2a)$$
or 
$$ H = e^{-xL}b^L \quad {\rm exponential}            \eqno(2b)$$
in our two choices. For the linear decay, the help is maximal for $L$ near
$x^{-1} - 1/\ln b$; for the exponential decay one has either $H \rightarrow \infty$
(percolation) or $H \rightarrow 0$ (no percolation) depending on whether 
$x$ is smaller or larger than ln $b$.

One may also look for a maximum of $H$ under the condition that the total
number $N$ of people, eq.(1), is constant. Then it is best to take 
$L = 1, \; b = N-1$, since then everybody is close to the truth, 

We doubt that these simple models and results are suitable ways to organise 
hospitals or other social organisations. 

\section{Scale-free networks}

Some but not all social relations may be better approximated by scale-free 
networks \cite{BA,simon,schnegg}. We use the Barab\'asi-Albert version, where
the number $n_k$ of network nodes having $k$ neighbours decays as $1/k^3$, Fig.
1. We start with $m$ fully connected nodes, and then each node which is newly 
added to the network selects $m$ existing nodes as ``bosses'', with a 
probability proportional to the number of neighbours the boss has at that time: 
preferential attachment.

While our information flow is directed, we do not need to take this direction
into account since no neighbour relations need to be stored. Instead, for each
newly added node we determine the shortest distance $L$ from the initial core
of $m$ nodes; this core has $L = 0$. Since now we have no longer a clear 
separation into layers, we assume that everybody except the core members 
helps the patients, with the fraction $f_i = \exp(-xL_i)$ (exponential decay) 
or $f_i = 1 - xL_i$ (linear decay). Then
$$H = \sum_{i=m+1}^N f_i \eqno(3)$$
is the total help. 
 
Fig.2 shows how the resulting help decreases with increasing loss fraction $x$.
It confirms, not surprisingly, that the help decreases with increasing 
information loss $x$ and increases with increasing network size $N$. (For 
fixed $N = 10^5$ and $2 \le m \le 7$ the help increases slightly with 
increasing $m$.) It is more realistic to include a cost associated with the 
network; thus we look at the modified help 
$$ H' = H - \lambda N, \quad  \lambda = 0.01 \quad . \eqno(4)$$
Fig.3 shows that this function has a maximum in the size range of interest,
except for small $x$ which would require even bigger networks. The 
logarithmic horizontal axis of Fig.3 corresponds to a linear axis in the
layer number $<\ell>$, since the latter varies as log $N$, Fig.4. Also
in the Cayley tree one can find such a maximum help if one subtracts 
$0.01 N$ from $H$, eqs.(1,2b)

\section{Conclusion}

For scale-free Barab\'asi-Albert networks and Cayley trees, we 
found a maximum in the desired help function at network sizes, which 
correspond to numbers of layers larger than two and smaller than ten,
a reasonable result.

We thank the Brazilian grants CNPq and FAPERJ for financial support and  
M. Izzo and F. Bagnoli for discussions.

\end{document}